\documentclass[preprint,12pt]{elsarticle}

\usepackage{amssymb}
\usepackage{amsmath}

\biboptions{sort&compress}

\journal{Scripta Materialia}

\usepackage{siunitx}

\usepackage{bm}

\usepackage{xcolor}%
\definecolor{LST}{rgb}{0.1, 0.6, 0.1}

\usepackage{ulem}

\newcommand{\figwidth}{0.8\columnwidth}

\begin{document}

\begin{frontmatter}


\title{Landau–Lifshitz–Bloch simulations of the magnetocaloric effect in continuous ferromagnetic–paramagnetic transitions}
\author[Valladolid]{Luis M. Moreno-Ramírez\corref{cor1}}%
\ead{luis.moreno.ramirez@uva.es}
\cortext[cor1]{Corresponding author}
\author[Valladolid]{Luis Sánchez-Tejerina\corref{cor2}}%
\ead{luis.sanchez-tejerina@uva.es}
\cortext[cor2]{Corresponding author}
\author[Valladolid]{Óscar Alejos}%
\affiliation[Valladolid]{organization={Dpto. Electricidad y Electrónica, Universidad de Valladolid, 47011 Valladolid, Spain}}%
\author[Sevilla]{Victorino Franco}%
\affiliation[Sevilla]{organization={Multidisciplinary Unit for Energy Science. Dpto. Física de la Materia Condensada, ICMSE-CSIC, Universidad de Sevilla, P.O. Box 1065, 41080-Sevilla, Spain}}
\author[Salamanca1,Salamanca2]{Víctor Raposo}
\affiliation[Salamanca1]{organization={Dpto. Física Aplicada, Universidad de Salamanca, 37008 Salamanca, Spain}}
\affiliation[Salamanca2]{organization={Unidad de Excelencia en Luz y Materia Estructuradas (LUMES), Universidad de Salamanca, Salamanca, Spain}}%

\begin{abstract}
The usefulness of modeling magnetocaloric materials expands from the understanding of their behavior to the prediction of new materials, playing a fundamental role in the optimization of their performance. In contrast with other areas of magnetic materials research, micromagnetic simulations of magnetocaloric materials are scarce due to the difficulty of modeling the material in the vicinity of the transition. To solve this limitation, we propose the use of micromagnetic simulations based on the Landau–Lifshitz–Bloch equation to study the magnetocaloric effect of a ferromagnetic material around its Curie transition. Following our proposed methodology, we obtain reliable isothermal entropy change curves for both monocrystalline and polycrystalline configurations, where we consider different anisotropic contributions. The robustness of the method was evaluated, yielding results that agreed with previous experimental and theoretical observations. Our study shows that micromagnetic simulations are a powerful tool for analyzing magnetocaloric materials with complex microstructures.
\end{abstract}




\end{frontmatter}

\section{Introduction}
\label{Introduction}
The increased energy consumption of domestic and industrial cooling systems, combined with existing energy losses, promotes the search for new refrigeration technologies that could offer high energy efficiency and low environmental impact \citep{Pezzutto2022, EuropeanUnion2025}. In this context, solid-state magnetic refrigeration has been extensively studied in recent decades \cite{Franco2018_progress, Kitanovski2020}. This technology exploits the magnetocaloric effect, which is quantified by the reversible adiabatic temperature change or the isothermal entropy change ($\Delta s_{\mathrm{iso}}$) when a magnetic material is subjected to a variable magnetic field \cite{Weiss1917_JPhysTheor}. A deep understanding of the link between magnetocaloric performance and material characteristics is key to the design of better magnetocaloric materials. This understanding can be achieved through several modeling approaches that focus on different length scales. At one extreme lie atomistic models, which are widely used for obtaining different structural and magnetic parameters of the materials \cite{Paudyal2006,PhysRevB.83.174403,Guillou2018,L.M.Moreno_horizons2019,Biswas2019}, although there are also works for reproducing or predicting the magnetocaloric performance \cite{Nbrega2005, Staunton2014, Comtesse2014, Bocarsly2017, Romero-Muiz2023, Almeida2024}. Unfortunately, these models are limited to a reduced number of atoms because of their time-consuming nature. At the other extreme are the macroscopic models based on different thermomagnetic models or equations of state, such as the Landau, Bean-Rodbell or Arrott-Noakes approaches \cite{PhysRevLett.19.786, Bean1962, Kuzmin2008}. Despite the demonstrated utility of both approaches, they cannot account for or oversimplify different features of real samples, such as magnetic domains or granular structures, although they significantly affect the magnetocaloric response \cite{Gottschall2017, Lai2021, Zhang2024, Daz-Garca2025, Prusty2025}. Neglecting this influence in the models have two deleterious effects: on the one hand, even if materials optimization significantly relies on engineering microstructure and domain structure, their influence cannot be properly described by conventional models. On the other hand, data analysis is also affected by this simplifications, in particular the field dependence of $\Delta s_{\mathrm{iso}}$, which are widely employed to probe the order of the transition and to determine the associated critical exponents \cite{Franco_2008_scalinglaws,FRANCO2010465,PhysRevB.81.224424,0022-3727-50-41-414004,Law2018_nature}.

Microscopic models appear to be an optimal alternative between these two extreme approaches, as they allow magnetic materials with complex microstructures to be replicated~\cite{Schrefl1994, Sepehri-Amin2013}. However, there have been limited attempts for magnetocaloric materials mainly due to the limitations of describing the material around its transition temperature. In this sense, the work in~\cite{Ohmer2020} stands out, where micromagnetic modeling based on the Landau-Lifshitz-Gilbert equation was employed up to temperatures close (but below) the Curie temperature of a ferromagnet while the behavior for higher temperatures was extrapolated by fitting previous micromagnetic results in the ferromagnetic range to the Arrott-Noakes equation of state~\cite{PhysRevLett.19.786}. Alternatively, in this work we propose the study of the magnetocaloric effect for both ferromagnetic (FM) and paramagnetic (PM) ranges of a ferromagnetic material employing micromagnetic simulations based on the Landau-Lifshitz-Bloch (LLB) equation~\cite{Garanin1997}. We reproduced the magnetocaloric effect in both monocrystalline and polycrystalline materials and evaluated the influence of sample geometry, magnetocrystalline anisotropy, and microstructure on $\Delta s_{\mathrm{iso}}$, paying special attention to the field dependence analysis, i.e., phenomenological universal curve and exponent $n$. The obtained results agree with both experimental and previous theoretical studies, showing that micromagnetic simulations are a powerful and useful tool for modeling realistic magnetocaloric materials. 


\section{Methods}
\label{Methods}
\begin{figure*}[t]
\centering
\includegraphics[width=0.9\textwidth]{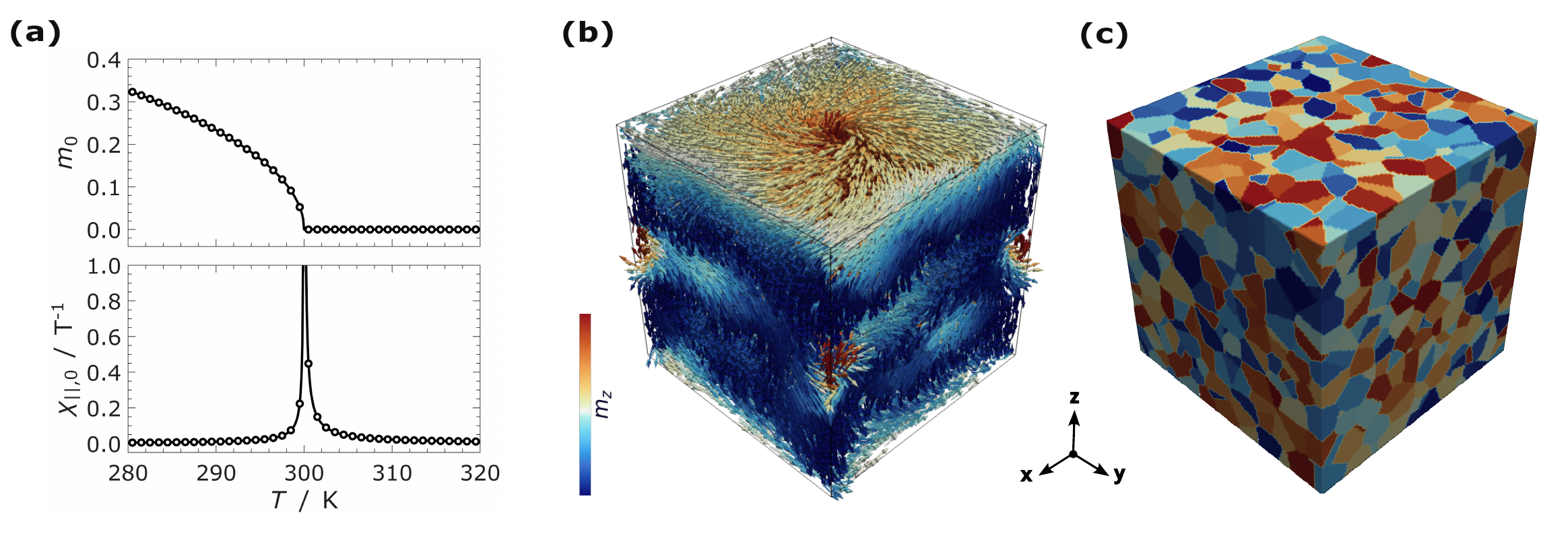} 
\caption{(a) Temperature dependence of LLB inputs: zero-field reduced magnetization (up) and reduced longitudinal susceptibility (down). (b) Equilibrium magnetization at $\qty{280}{\kelvin}$ and $\qty{0.001}{\tesla}$ for a monocrystal composed of a $128\times128\times128$ cells. (c) Polycrystalline sample with microstructure generated by Voronoi tesselation (each grain region is colored to distinguish it).}
\label{fig:esquema}
\end{figure*}

We implemented into a customized micromagnetic software based on MuMax$^{3}$~\cite{Vansteenkiste2014} the classical LLB equation to obtain the total equilibrium magnetization ($\boldsymbol{M}$) of a ferromagnetic material around its Curie temperature ($T_{\mathrm{C}}$) for different applied fields ($\boldsymbol{H}$). Further details of the simulations can be found in~\cite{Moretti2017} and in section S1 of the supplementary material. As temperature-dependent material inputs for the simulations, it has to be introduced for each micromagnetic cell the magnitude of the reduced magnetization at zero applied field ($m_{0}$) and the longitudinal reduced magnetic susceptibility for zero applied field $\left( \chi_{\parallel,\,0}\right)$. We employed the Brillouin theory in the classical limit for both inputs \cite{Coey2010}:
\begin{gather}
m_{0} = \mathcal{L}(x)\;\label{eq:1},\\
\chi_{\parallel,\, 0} = \frac{\mu\mathcal{L}'(x)}{k_{B}T\left(1-3T_{\mathrm{C}}\mathcal{L}'(x)/T\right)} \; \label{eq:2},
\end{gather}
where $\mathcal{L}$ and $\mathcal{L}'$ are the Langevin function and its derivative, respectively, $x=3m_{0}T_{\mathrm{C}}/T$, $\mu$ the atomic magnetic moment, and $k_{B}$ the Boltzmann constant. The employed inputs with chosen material parameters (described in this section) are depicted in Figure \ref{fig:esquema} (a). Other inputs based on measurement or ab-initio calculation can also be employed without compromising the validity of our approach. This study aims to show that full micromagnetic simulations can be used to model magnetocaloric materials, rather than focusing on a specific material. For the simulations, cubic micromagnetic cells with $\qty{4}{\nano\metre}$ edges were employed for different sample geometries. This cell size is chosen to properly characterize the changes in magnetization between magnetic domains according to the selected material parameters. As an example, the equilibrium magnetization solutions for $128\times128\times128$ cells is shown in panel (b) of Figure \ref{fig:esquema}. We considered $\mu=1 \, \mu_{\mathrm{B}}$ (where $\mu_{\mathrm{B}}$ is the Bohr magneton), $T_{\mathrm{C}}=\qty{300}{\kelvin}$, saturation magnetization ($M_{\mathrm{sat}}$) of $\qty{407}{\kilo \ampere \per \meter}$ ($\qty{80}{\ampere\square\meter\per\kilogram}$), and exchange stiffness at $\qty{0}{\kelvin}$ ($A_{0}$) of $\qty{1.4}{\pico \joule \per \meter}$. We introduced an uniaxial magnetocrystalline anisotropy following a classical approach for the first order anisotropy constant ($K_{\mathrm{u}}$) to be proportional to $m_{0}^{2}$ in the vicinity of the transition \cite{Landau1984, Coey2010}, being $\qty{0.2}{\mega \joule \per \cubic \meter}$ at $\qty{280}{\kelvin}$ \cite{Josten2025}. The selected parameters are similar to those of the AlFe$_{2}$B$_{2}$ alloy, an interesting rare-earth-free second-order magnetocaloric material for room temperature applications \cite{Tan2013,Lewis2015,Lamichhane2018,Beckmann2023}. The different granular microstructures of the polycrystalline samples were generated by the Voronoi tesselation method implemented in MuMax$^{3}$ \cite{Leliaert2014}. An example of the generated microstructures is depicted in panel (c) of Figure \ref{fig:esquema}.\\

To evaluate the magnetocaloric effect, $\Delta s_{\mathrm{iso}}$ is calculated from the total magnetization through the Maxwell relations following a discrete approximation to:
\begin{equation}\label{eq:3}
\Delta s_{\mathrm{iso}}=\mu_{0} \int_{\boldsymbol{0}}^{\boldsymbol{H_{\mathrm{f}}}}\left( \frac{\partial \boldsymbol{M}}{\partial T} \right)_{\boldsymbol{H}} \cdot \text{d} \boldsymbol{H},
\end{equation}
where $\boldsymbol{H_{\mathrm{f}}}$ is the final applied magnetic field \cite{doi:10.1063/1.370767}. The scaling behavior of $\Delta s_{\mathrm{iso}}$ is tested by the phenomenological universal curve \cite{Franco_2008_scalinglaws,FRANCO2010465}. This curve is constructed by rescaling both $\Delta s_{\mathrm{iso}}$ and temperature axis for different magnetic field changes. First, $\Delta s_{\mathrm{iso}}$ is normalized by its peak value for each field ($\Delta s_{\mathrm{iso}}^{\mathrm{pk}}$). Second, for rescaling the temperature axis, the reference temperatures ($T_{\mathrm{r}}$) are defined as the temperature for which the normalized value of $\Delta s_{\mathrm{iso}}$ is the same in both paramagnetic and ferromagnetic ranges (0.7 in this study). Once $T_{\mathrm{r}}$ is obtained for each field, the temperature axis is rescaled as $(T-T_{\mathrm{pk}})/(T_{\mathrm{r}}-T_{\mathrm{pk}})$ where $T_{\mathrm{pk}}$ is the peak temperature. The field dependence exponent $n$ is calculated following a discrete approximation to:
\begin{equation}\label{eq:4}
    n=\frac{\partial \ln (\vert \Delta s_{\text{iso}} \vert / \vert \Delta s_{\text{iso}}(1 \, \mathrm{T}) \vert)}{\partial \ln ( \Delta H / (1 \, \mathrm{T}))},
\end{equation}
where $\Delta H$ is the applied magnetic field change module. 

The simulations were conducted using external conditions similar to the experimental ones to calculate \eqref{eq:3} and \eqref{eq:4}. A minimal temperature step of 1 K around $T_{\mathrm{C}}$ is selected. For the magnetic field, we simulated up to 40 fields from 0.001 T to 2 T following a quadratic progression. 

\section{\label{Results} Results and discussion}

\begin{figure}[t]
\centering
\includegraphics[width=\figwidth]{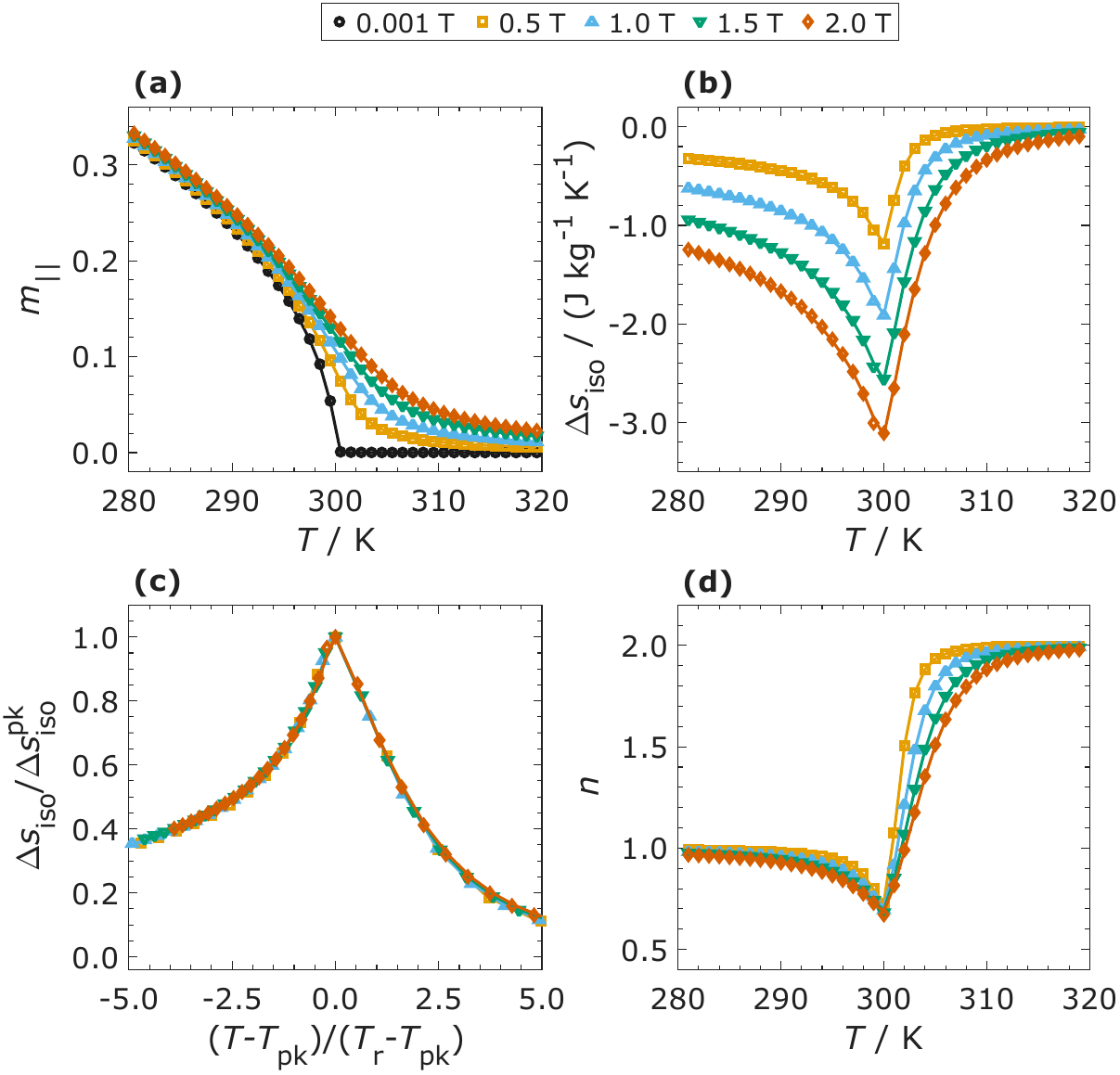}
\caption{Temperature dependence of the (a) longitudinal magnetization, (b) isothermal entropy change, (c) phenomenological universal curve, and (c) exponent $n$ for different applied fields and field changes. The hollow symbols correspond to the micromagnetic results, while the solid lines corresponds to the Brillouin solutions. Most symbols appear filled as they are superimposed with solid lines.}
\label{fig:Figure_homogeneo}
\end{figure}

Initially, a monocrystalline material was reproduced without considering any anisotropy and resulting in a homogeneously magnetized magnetic material. In this way, the simulation results can be compared with those from macroscopic models with single-domain magnetization. For this comparison, we selected the Brillouin theory in the classical limit to ensure consistency with the employed inputs and because it has proven highly useful in the modeling of magnetocaloric materials \cite{DIAMANTOPOULOS2025359}. The results of the simulations are depicted as hollow symbols in Figure~\ref{fig:Figure_homogeneo} while the corresponding results from the Brillouin theory are plotted as solid lines. A good agreement was observed between the simulations and Brillouin results for all the explored magnitudes. Additionally, despite using zero-field inputs, the micromagnetic simulations reproduce all the magnitudes quite well, even for 2 T. For the magnetization component parallel to the applied field ($m_{||}$), the deviations with respect to the Brillouin theory are below $0.1 \, \%$ in the temperature and field ranges explored (Figure~\ref{fig:Figure_homogeneo} (a)). For $\Delta s_{\mathrm{iso}}$, the deviations are found to be below $1.9 \, \%$, being more significant around $T_{\mathrm{C}}$ and for the larger applied field changes (Figure~\ref{fig:Figure_homogeneo} (b)). These minor deviations have a minor influence on subsequent field dependence analysis, as can be observed from the excellent collapse of the universal construction (Figure~\ref{fig:Figure_homogeneo} (c)). Additionally, the exponent $n$ magnitude showed the expected behavior for a Curie transition (Figure~\ref{fig:Figure_homogeneo} (d)): a value equal to 1 in the ferromagnetic range, 2 in the paramagnetic one, and 2/3 at $T_{\mathrm{C}}$. From the simulations $n(T_{\mathrm{C}})=0.696$, deviating a $4.5 \, \%$ of the theoretical value. This deviation is moderate considering that to calculate the exponent $n$: 1) the initial magnetization data is derived twice (eqs. \eqref{eq:3} and \eqref{eq:4}) and 2) a discrete approach is followed. In this sense, similar deviations are found for the Brillouin calculations. These findings demonstrate that the simulations successfully replicate the magnetocaloric response. From now on, the comparison with Brillouin theory cannot be established as we will incorporate different contributions to the energy density which leads to multi-domain regimes, thus discarding the sample description as a whole.

\begin{figure}[t]
\centering
\includegraphics[width=\figwidth]{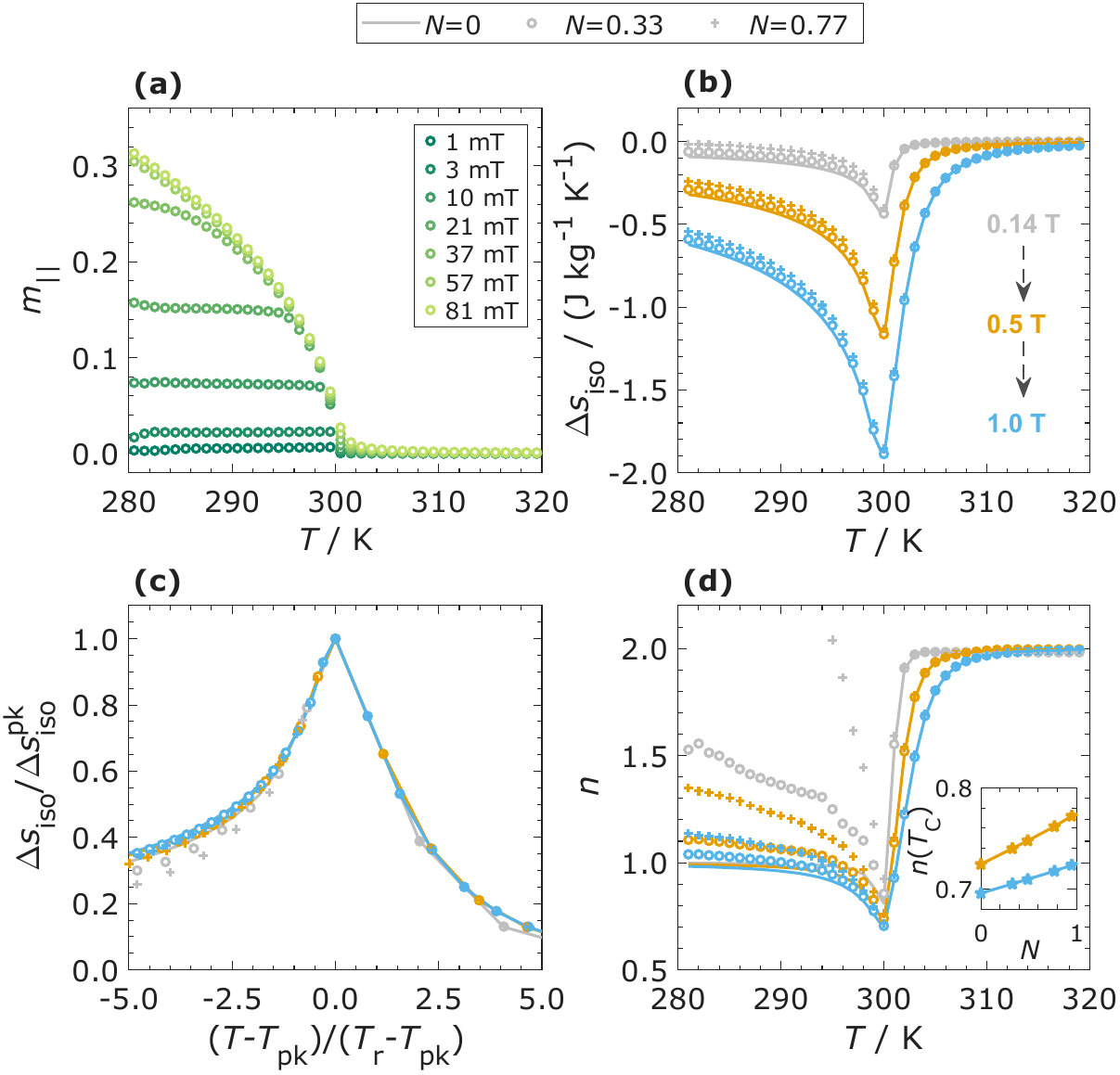}
\caption{(a) Temperature dependence of the magnetization for low applied fields for a sample with cubic geometry. Temperature dependence of the (b) isothermal entropy change, (c) phenomenological universal curve, and (d) exponent $n$ for different samples and orientations (labeled by an effective demagnetizing factor $N$). Panel (d) inset: exponent $n$ at $T_{\mathrm{C}}$ as a function of $N$ for 0.5 T and 1.0 T. The hollow circles are superimposed with symbols and appear as filled circles in the paramagnetic range.}
\label{fig:efecto-desimanador}
\end{figure}

We next examined the demagnetizing field effect on the magnetocaloric response. For that, we simulated a monocrystalline material with different sample geometries and field orientations. The micromagnetic solver allows the introduction of the demagnetizing field by using a discrete convolution of the magnetization with a demagnetizing kernel \cite{Vansteenkiste2014}. From the simulation results, we calculated the corresponding effective demagnetizing factor ($N$) associated to the relevant axis to serve merely as a label. The corresponding geometry is tabulated in section S2 of the supplementary material.  As an example, the magnetization data for low applied fields for a cubic sample composed of $128\times128\times128$ cells (label $N=0.33$) are shown in panel (a) of the Figure~\ref{fig:efecto-desimanador}. It is observed that the demagnetizing field reduces the total magnetization of the sample in the FM range compared to the previous values shown in Figure~\ref{fig:Figure_homogeneo}. This effect is overcome as the applied field increases up to $\sim$ 80 $\unit{\milli \tesla}$ in the explored temperature range, being slightly above the demagnetizing one ($\sim \qty{55}{\milli \tesla}$ at 280~$\unit{\kelvin}$). The isothermal entropy change, universal construction, and exponent $n$ for different sample geometries and applied field orientations (labeled by $N$) are shown in panels (b), (c), and (d) of Figure~\ref{fig:efecto-desimanador}, respectively.

It is evidenced that $\Delta s_{\mathrm{iso}}$ values are reduced by the effect of the demagnetizing field, showing an almost constant reduction in the ferromagnetic range that vanishes slightly above $T_{\mathrm{C}}$. In addition, the larger the demagnetizing factor, the larger the deviations, as expected. These deviations saturate for fields slightly above the demagnetizing field, becoming almost field independent after that. For applied fields with technological interest, such as 1 $\unit{\tesla}$, its effect is moderate (e.g., deviations of the peak are -2.9 $\%$ for the case labeled as $N=0.77$). In addition, they degrade the phenomenological collapse for low applied field changes and low temperatures but, if the analysis is carried out for fields clearly above the demagnetizing field, its effect is negligible. Finally, the demagnetizing field significantly overestimated the exponent $n$ for $T\leq T_{\mathrm{C}}$. Similar to the previous results, the deviations increase as the demagnetizing factor increases and the temperature decreases. The magnetic field change strength tends to diminish the deviations, although their effects are still significant for 1 $\unit{\tesla}$. There is also a significant effect at $T_{\mathrm{C}}$, where the exponent $n$ is related to the material critical exponents. A linear increasing trend between $n(T_{\mathrm{C}})$ and $N$ is observed (inset of panel (d) of Figure \ref{fig:efecto-desimanador}). The data deviate from the expected theoretical values as the demagnetizing factor increases, which can lead to erroneous interpretations of the critical exponents. The insights obtained from the simulations agree with previous theoretical and experimental results \cite{doi:10.1063/1.3067463, doi:10.1063/1.4885110,Moreno-Ramirez2015_jallcom}, highlighting the importance of reducing the demagnetizing fields to obtain accurate physical interpretation of the magnetocaloric effect.

\begin{figure}[t]
\centering
\includegraphics[width=\figwidth]{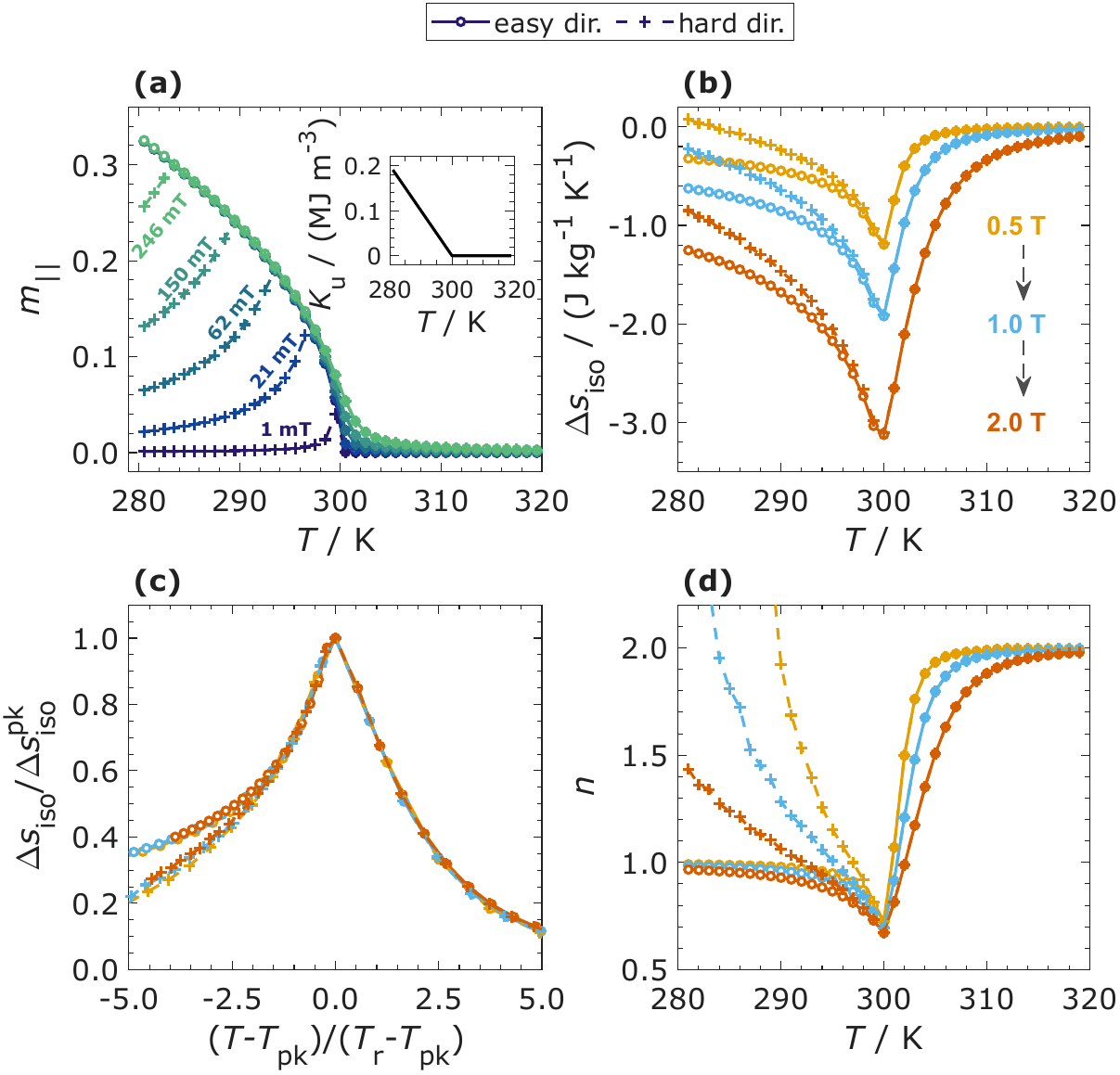}
\caption{(a) Temperature dependence of the magnetization for low applied fields along easy and hard directions for a sample with temperature-dependent uniaxial anisotropy (inset). Calculated (b) isothermal entropy change, (c) phenomenological construction, and (d) exponent $n$ for different applied field changes along easy and hard axis. The hollow circles are superimposed with symbols and appear as filled circles in the paramagnetic range.}
\label{fig:Figure_anisotropia}
\end{figure}

In the following we included the magnetocrystalline anisotropy as described previously. The temperature dependence of $K_{\mathrm{u}}$ is plotted in the inset of panel (a) of Figure \ref{fig:Figure_anisotropia}. A square thin sheet plate of $512\times512\times8$ cells with in-plane uniaxial anisotropy axis having both easy and hard orientations lying in the sample plane is simulated. Magnetization, isothermal entropy change, phenomenological universal curve, and exponent $n$ with the applied field along both easy and hard orientations are shown in Figure \ref{fig:Figure_anisotropia}. The magnetization along the easy orientation coincides with the total magnetization, whereas the values for the hard orientation are considerably reduced for low fields, showing an increasing trend with temperature until the anisotropy field is overcome. The larger the applied field, the lower the temperature required to saturate the magnetization. When the anisotropy field is overcome, the magnetization is saturated for both easy and hard axes. For $\Delta s_{\mathrm{iso}}$ two differentiated responses are observed according to the orientation, being smaller for the hard orientation. These differences vanish at $T_{\mathrm{C}}$. For the universal construction, two different branches associated with each orientation are observed, showing a poorer collapse along the hard direction than that along the easy one. This behavior is found experimentally as shown in \cite{Liu2020}. Comparing to AlFe$_{2}$B$_{2}$ \cite{Barua2018, Tran2022}, the anisotropic behavior seems to be extended up to the paramagnetic range in contrast to the material parameters here considered. However, the experimentally observed temperature and field dependence of $K_{\mathrm{u}}$ can be straightforwardly included in our approach if available. For the exponent $n$, applying the field along the hard axis causes strong deviations in the ferromagnetic range from the theoretical value of 1, similar to the effects of shape anisotropy. Alternatively, the anisotropic magnetocaloric behavior can be exploited in devices with rotating fields \cite{Balli2014,Almeida2024}. Our approach also account for this behavior as shown in section S3 of the supplementary material.

\begin{figure}[t]
\centering
\includegraphics[width=\figwidth]{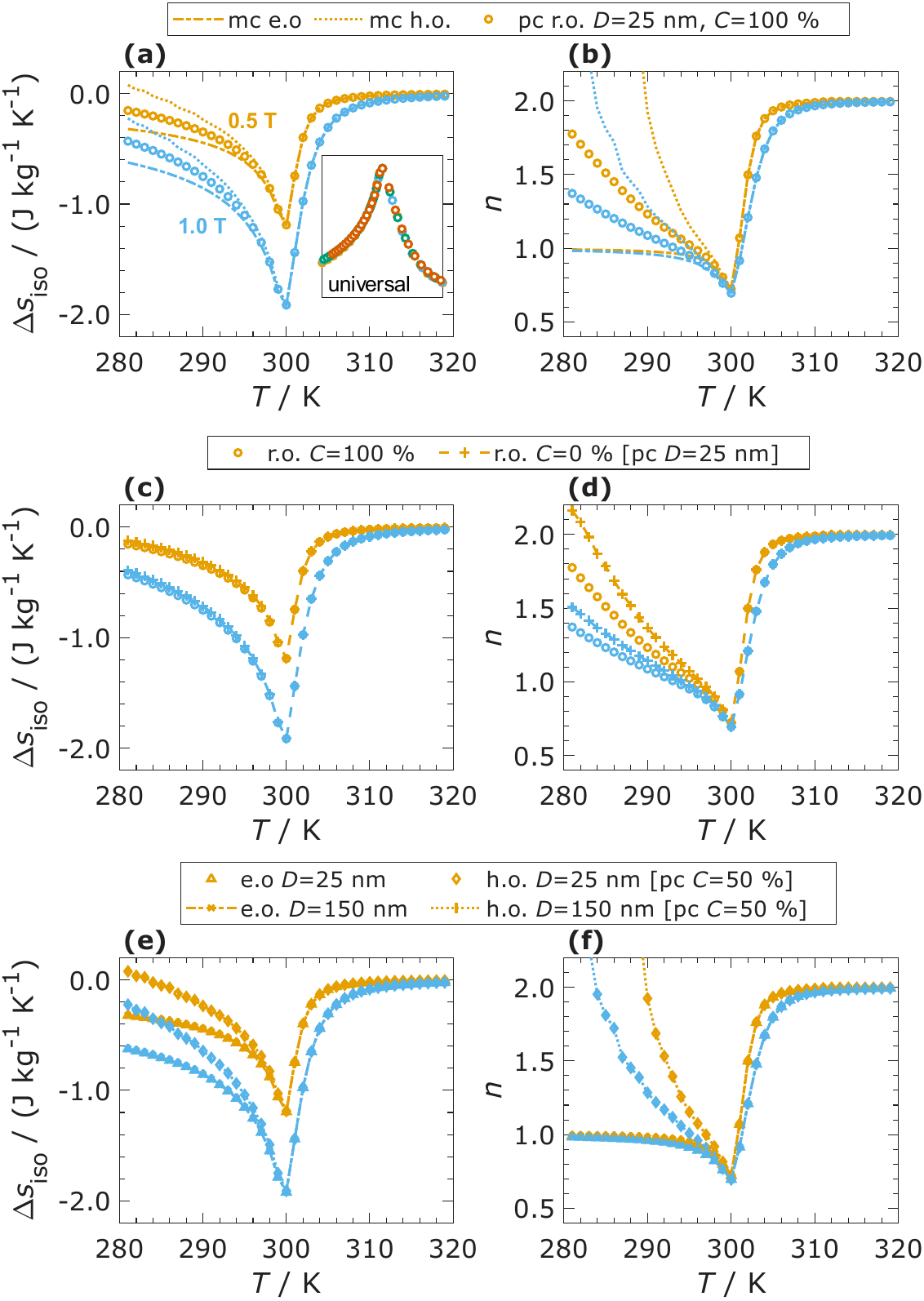}
\caption{Isothermal entropy change (left) and exponent $n$ (right) magnitudes for: 1) polycrystal (pc) with randomly orientated grains and monocrystal (mc) for both easy (e.o.) and hard (h.o.) orientations [(a) and (b)], 2) polycrystal with randomly orientated grains with different exchange coupling between grains, $C$ [(c) and (d)] and 3) oriented polycrystal with different mean grain size, $D$ [(e) and (f)]. Corresponding phenomenological construction of panel (a) data is shown in as an inset (same scale and units as previous plots).}
\label{fig:Figure_granos}
\end{figure}

Finally, granular microstructures with different crystalline orientation distributions were generated using Voronoi tessellation. For all cases, a thin square sheet is used to avoid in-plane demagnetizing effects. The magnetocaloric effect is simulated for a randomly oriented polycrystal with a mean grain size ($D$) of $\qty{25}{\nano\meter}$ and $100 \,\%$ exchange coupled grains ($C=100 \, \%$). $\Delta s_{\mathrm{iso}}$ and exponent $n$ are depicted in panels (a) and (b) of Figure~\ref{fig:Figure_granos}, respectively, together with the previously characterized monocrystal response. As expected, and experimentally observed, the polycrystal response for both magnitudes is the average of the easy and hard responses of the monocrystal~\cite{Fries10.1063/1.4971839}. The universal construction is depicted in the inset of panel (a) (the representation range is the same as in all previous cases). A good collapse can be observed, which is an interesting result despite the double universal curve observed for the uniaxial monocrystal. This demonstrates the robustness of the phenomenological construction. Conversely, for exponent $n$, it is observed that anisotropy shifts the values from the expected ones although this effect is mitigated in comparison to the monocrystalline case. We have also analyzed the influence of exchange coupling between grains by reducing it from fully coupled to decoupled, as shown in panels (c) and (d) of Figure~\ref{fig:Figure_granos}. Decoupling the grains causes a minor effect on $\Delta s_{\mathrm{iso}}$. Additionally, its effect on the exponent $n$ is more significant, increasing the values in the FM range. The lower the field change, the larger the effect. Finally, we decided to align all the grains with the same orientation to check the influence of the mean grain size without the undesired effect of not fully random distributed orientations in the case of large grains. The obtained magnitudes for grain-oriented distributions with $\qty{25}{\nano\meter}$ and $\qty{150}{\nano\meter}$ mean sizes are illustrated in panels (e) and (f) of the Figure~\ref{fig:Figure_granos}. No significant differences were observed for either the orientation or applied field changes. These results indicate that grain boundary effects are minimal in magnetocaloric materials that undergo second-order magnetic transitions. In comparison to~\cite{Ohmer2020}, we were not considering nonmagnetic grain boundaries; therefore, the mass of magnetocaloric material is not reduced in our test.

\section{Conclusions}
\label{Conclusions}
Using micromagnetic simulations based on the Landau-Lifshitz-Bloch equation, we were able to reproduce the magnetocaloric response associated with a Curie transition. Using generic parameters similar to those of AlFe$_{2}$B$_{2}$, a rare-earth-free magnet with a notable magnetocaloric effect, we calculated the magnetocaloric response in both monocrystal and polycrystal configurations, being the latter case challenging to analyze through macroscopic or atomic approaches, which exemplifies the virtues of our proposed methodology. The effect of the shape and magnetocrystalline anisotropy on the magnetocaloric effect is analyzed, paying special attention to the field-dependence behavior through the universal curve and exponent $n$. The results obtained are in agreement with previous experimental or theoretical approaches, demonstrating the capability of micromagnetic simulations to characterize second-order magnetocaloric materials with different microstructures. Furthermore, this work provides an opportunity to further improve the model to characterize magnetocaloric materials.

\clearpage

\section*{Acknowledgments}
We acknowledge funding from Projects PID2023-150853NB-C31 and PID2023-146047OBI00 funded by MICIU/AEI/10.13039/501100011033 and FEDER, and project Magccine funded by the European Union and the European Innovation Council.

\bibliographystyle{elsarticle-num} 
\bibliography{export}
\end{document}


\begin{frontmatter}

\title{Supplementary Information: Landau–Lifshitz–Bloch simulations of the magnetocaloric effect in continuous ferromagnetic–paramagnetic transitions}

\author[Valladolid]{Luis M. Moreno-Ramírez}%
\author[Valladolid]{Luis Sánchez-Tejerina}%
\author[Valladolid]{Óscar Alejos}%
\affiliation[Valladolid]{organization={Dpto. Electricidad y Electrónica, Universidad de Valladolid, 47011 Valladolid, Spain}}%
\author[Sevilla]{Victorino Franco}%
\affiliation[Sevilla]{organization={Multidisciplinary Unit for Energy Science. Dpto. Física de la Materia Condensada, ICMSE-CSIC, Universidad de Sevilla, P.O. Box 1065, 41080-Sevilla, Spain}}%
\author[Salamanca1,Salamanca2]{Víctor Raposo}
\affiliation[Salamanca1]{organization={Dpto. Física Aplicada, Universidad de Salamanca, 37008 Salamanca, Spain}}%
\affiliation[Salamanca2]{organization={Unidad de Excelencia en Luz y Materia Estructuradas (LUMES), Universidad de Salamanca, Salamanca, Spain}}%

\end{frontmatter}                     
\thispagestyle{plain}
\section{\label{LLB} Landau-Lifshitz-Bloch equation}

The classical Landau–Lifshitz–Bloch (LLB) equation is given by \cite{PhysRevB.55.3050, PhysRevB.74.094436,Moretti2017}:
\begin{equation}
\begin{split}
\frac{\mathrm{d}\boldsymbol{m}}{\mathrm{d}t}&=-\gamma_{0} \left( \boldsymbol{m} \times \boldsymbol{H_{\mathrm{eff}}} \right) +\frac{\gamma_{0}\alpha_{\parallel}}{m^{2}}\left( \boldsymbol{m} \cdot \boldsymbol{H_{\mathrm{eff}}}\right)\boldsymbol{m}\\
&{\quad} -\frac{\gamma_{0}\alpha_{\perp}}{m^{2}}\left( \boldsymbol{m} \times \left( \boldsymbol{m} \times \boldsymbol{H_{\mathrm{eff}}} \right) \right),
\end{split}
\end{equation}
where $\boldsymbol{m}$ is the reduced magnetization (i.e. $\boldsymbol{M}/M_{\mathrm{sat}}$ being $M_{\mathrm{sat}}$ the saturation magnetization at $\qty{0}{\kelvin}$), $\gamma_{0}$ the gyromagnetic ratio (which has a positive value), $\alpha_{\parallel}$ and $\alpha_{\perp}$ the longitudinal and transversal damping parameters, respectively, and  $\boldsymbol{H_{\mathrm{eff}}}$ the effective magnetic field. This equation relates the temporal evolution of $\boldsymbol{m}$ with a precessional term (first term on the right-hand side), and two damping terms that are longitudinal and transversal to the effective field (second and third terms on the right-hand side, respectively). In this work, we are interested in the equilibrium solution and, therefore, the damping constants have been selected as 0.1 for speeding up the simulations. The effective field is considered as a sum of different contributions: the applied field ($\boldsymbol{H}$), the anisotropy field ($\boldsymbol{H_{\mathrm{ani}}}$), the demagnetizing field ($\boldsymbol{H_{\mathrm{dem}}}$), and the exchange field ($\boldsymbol{H_{\mathrm{exc}}}$). The exchange one is considered as:
\begin{equation}
\boldsymbol{H_{\mathrm{exc}}} =\frac{2A}{\mu_{0}M_{\mathrm{sat}}m_{0}^{2}}\nabla^{2}\boldsymbol{m}+\begin{cases}
    \frac{1}{2\chi_{\parallel}}\left( 1-\frac{m^{2}}{m_{0}^{2}} \right) \boldsymbol{m} & T \lesssim T_{C},\\
    {-\frac{1}{\chi_{\parallel}}}\left( 1+\frac{3}{5}\frac{T_{C}}{T-T_{C}} \right) \boldsymbol{m} & T \gtrsim T_{C},
\end{cases}
\label{eq:2}
\end{equation}
where $A$ is the exchange stiffness, $m_{0}$ is the reduced equilibrium magnetization at zero field, and $\chi_{\parallel}$ is the reduced longitudinal susceptibility at zero field $\left( \chi_{\parallel,\,0}=\left(\frac{\partial m}{\partial \mu_{0}H}\right)_{H \rightarrow 0} \right)$, with SI units of $\unit{\per \tesla}$. The last term on the right-hand side in equation \ref{eq:2} is added to account for the longitudinal exchange field, which promotes the module of the reduced magnetization towards its equilibrium value.  The equilibrium magnetization as well as the longitudinal susceptibility should be introduced based on the characteristics of the material. Following a mean-field approximation, the temperature dependence of $A$ is given by $A_{0}m_{0}^2$, where $A_{0}$ is the exchange stiffness at $\qty{0}{\kelvin}$.

\section{\label{geometries} Demagnetizing factors of tested geometries}
The different tested geometries (rectangular parallelepiped with dimensions $L_{x}\times L_{y}\times L_{z}$) and their corresponding label based on their effective demagnetizing factor along the principal directions are listed in the Table~\ref{tab:demagnetizing factors}. The demagnetizing factors are calculated from the low applied field behavior.

\begin{table}[h!]
    \centering
    \begin{tabular}{|c|c|c|c|}
    \hline
        Sample geometry & $N_{x}$ & $N_{y}$ & $N_{z}$\\
        \hline
        $\qty{512}{\milli\meter}\times\qty{512}{\milli\meter}\times\qty{512}{\milli\meter}$ & \multicolumn{3}{c|}{$0.3332$}\\
        \hline
        $\qty{1024}{\milli\meter}\times\qty{1024}{\milli\meter}\times\qty{128}{\milli\meter}$ & \multicolumn{2}{c|}{$0.1132$} & $0.7733$\\
        \hline
        $\qty{2048}{\milli\meter}\times\qty{2048}{\milli\meter}\times\qty{32}{\milli\meter}$ & \multicolumn{2}{c|}{$0.0242$} & $0.9510$\\
        \hline
        $\qty{128}{\milli\meter} \times \qty{128}{\milli\meter}\times\qty{4096}{\milli\meter}$ & \multicolumn{2}{c|}{$0.4926$} & $0.0145$\\
        \hline
    \end{tabular}
    \caption{Demagnetizing factors for different rectangular parallelepipeds.}
    \label{tab:demagnetizing factors}
\end{table}

\section{\label{rotating} Rotating response}
\begin{figure}
\includegraphics[width=0.98\columnwidth]{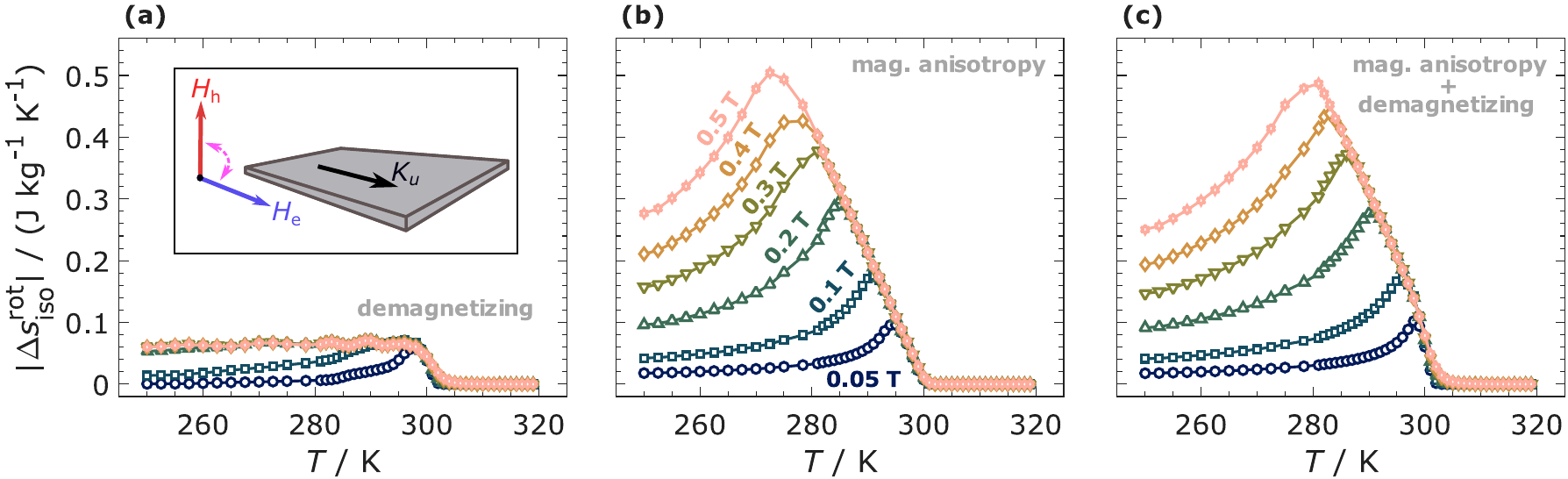}
\caption{Absolute rotating isothermal entropy change for a thin square sheet ($\qty{2048}{\nano\meter}\times\qty{2048}{\nano\meter}\times\qty{32}{\nano\meter}$) with in-plane uniaxial anisotropy for in-plane to out-of-plane field rotation (inset) while considering (a) the demagnetizing contribution, (b) the magnetocrystalline contribution and (c) both demagnetizing and magnetocrystalline contributions (c). The color code for the applied field is the same for the three panels.}
\label{fig:Figure_rotatorio}
\end{figure}
The anisotropic magnetocaloric response induced by both shape and magnetocrystalline anisotropy can be exploited to obtain a significant rotating magnetocaloric effect \cite{Balli2014,Almeida2024}. This rotating response ($\Delta s_{\mathrm{iso}}^{\mathrm{rot}}$) is calculated from the difference responses along two perpendicular orientations and, therefore, it can also be obtained from the proposed approach. We selected a thin square sheet with temperature-dependent uniaxial anisotropy (as described in the methods section) with a rotating applied field whose axis of rotation is contained in the plane of the sample, to maximize the influence of shape anisotropy. An schematic representation is shown in the inset of Figure~\ref{fig:Figure_rotatorio}. To characterize the response caused  exclusively by shape anisotropy, the magnetocrystalline anisotropy constant was set to zero. The rotating magnetocaloric response for the shape anisotropy is shown in Figure~\ref{fig:Figure_rotatorio} (a). The effective demagnetizing factors for the selected orientations are $0.0242$ and $0.9510$ for in-plane and out-of-plane orientations, respectively, which are close to the maximum theoretical range of 0 and 1. The rotating magnetocaloric response peaks slightly below $T_{\mathrm{C}}$ for low fields. As the magnetic field strength increases, it evolves to a plateau-like behavior for temperatures below $T_{\mathrm{C}}$. The response saturates in the explored range at approximately $\qty{0.2}{\tesla}$, showing a plateau value of $\qty{0.07}{\joule\per\kilogram\per\kelvin}$. This value is approximately $\qty{12}{\percent}$ of the conventional response for $\qty{0.2}{\tesla}$. The contribution of the magnetocrystalline anisotropy without considering demagnetizing effects is shown in Figure~\ref{fig:Figure_rotatorio} (b). The maximum rotating isothermal entropy change increases with the field strength and shifts to lower temperatures at a rate of about $\qty{-40}{\kelvin\per\tesla}$, considerably increasing the peak width. Figure~\ref{fig:Figure_rotatorio} (c) shows the calculated rotating response considering both demagnetizing and magnetocrystalline contributions. The demagnetizing field does not significantly affect the values achieved by the magnetocrystalline anisotropy and slightly reduces the shift of the peak to lower temperatures down to $\qty{-35}{\kelvin\per\tesla}$. The rotating response is approximately $\qty{40}{\percent}$ of the conventional response for $\qty{1}{\tesla}$, in agreement with experimental results \cite{Barua2018}.
\bibliographystyle{plain}
\bibliography{export2}